\documentclass[fleqn,usenatbib]{mnras}

\usepackage[T1]{fontenc}
   \usepackage{amsmath,amssymb,amsbsy,amsfonts}
   \usepackage{hyperref}
\DeclareRobustCommand{\VAN}[3]{#2}
\let\VANthebibliography\thebibliography
\def\thebibliography{\DeclareRobustCommand{\VAN}[3]{##3}\VANthebibliography}

\usepackage[draft]{graphicx}	% Including figure files
   \usepackage{amsmath,amssymb,amsbsy,amsfonts}
\usepackage{comment}
\usepackage{hyperref,graphicx}    
\usepackage{epstopdf}

\newcommand*{\figref}[2][]{%
  \hyperref[{#2}]{%
    Figure~\ref*{#2}%
    \ifx\\#1\\%
    \else
      \,#1%
    \fi
  }%
}
\usepackage{tikz}
\usetikzlibrary{arrows,matrix}

\title[Quasi-periodic gamma-ray oscillation in Sgr~A\(^*\)]{A
78- and 96- minute quasi-periodic gamma-ray oscillations in
Sagittarius\,A\(^*\)}

\author[G. Magallanes-Guij\'on \& S. Mendoza]{
Gustavo Magallanes-Guij\'on\thanks{E-mail:
gustavo.magallanes.guijon@ciencias.unam.mx (GM)}
and Sergio Mendoza\thanks{
E-mail: sergio@astro.unam.mx (SM)}
\\
Instituto de Astronom\'{\i}a, Universidad Nacional
                 Aut\'onoma de M\'exico, AP 70-264, Ciudad de M\'exico 04510,
                 M\'exico}

\date{Accepted XXX. Received YYY; in original form ZZZ}

\pubyear{2025}

\begin{document}
\label{firstpage}
\pagerange{\pageref{firstpage}--\pageref{lastpage}}
\maketitle

\begin{abstract}
Using publicly available \(\gamma\)-rays data from the Fermi satellite
and the Fermitools package, we constructed 867-day light curves of
Sagittarius A* (Sgr~A*) to search for potential periodicities from 2008
to 2024. By applying statistical and computational inference methods,
we used periodograms, a window function, unsupervised machine
learning (\textit{K}-Means), the Markov Chain Monte Carlo method,
noise colour analysis, and phase-folding techniques. Additionally, the
datasets were fitted with a Jacobi elliptical function using a likelihood
approach. In total, 1,463,040 minutes were analysed.  This analysis revealed
a significant period of enhanced activity lasting \(\sim\)~78 minutes,
followed by a quiescent phase of about \(\sim\)~18 minutes, resulting
in a total quasi-periodic oscillation (QPO) of around 96 minutes.
\end{abstract}

\begin{keywords}
% gamma-ray astronomy -- quasi-periodic oscillations -- Sagittarius A* 
gamma-rays: general -- gamma-rays: galaxies
\end{keywords}

%%%%%%%%%%%%%%%%%%%%%%%%%%%%%%%%%%%%%%%%%%%%%%%%%%

%%%%%%%%%%%%%%%%% BODY OF PAPER %%%%%%%%%%%%%%%%%%

%%%%%%%%%%%%%%%%%%%%%%%%%%%%%%%%%%%%%%%%%%%%%%%%%%

\section{Introduction}
\label{introduction}

\noindent
Sagittarius~A* (Sgr~A*) is a compact radio source situated at the Galactic
centre at a distance of \(8.2\)~kpc from Earth~\citep{Abuter2019A}
with a R.A. $17^{\mathrm{h}} 45^{\mathrm{m}} 40.05^{\mathrm{s}}$
and Dec. \(-29^\circ\) \(00'\) \(27.9''\).  It is an ideal candidate for a
supermassive black hole (SMBH) of a mass \(\sim 4.1 \times \mathrm{10}^{6}
\ \mathrm{M_{\odot}}\)~\citep{Boehle2016ApJ, Genzel2010, Ghez2008ApJ}.

%%===============================%%
%%==== B E G I N  F I G U R E ===%%
%%===============================%%
\begin{figure*}%[h!]
\centering 
\includegraphics[width=\textwidth]{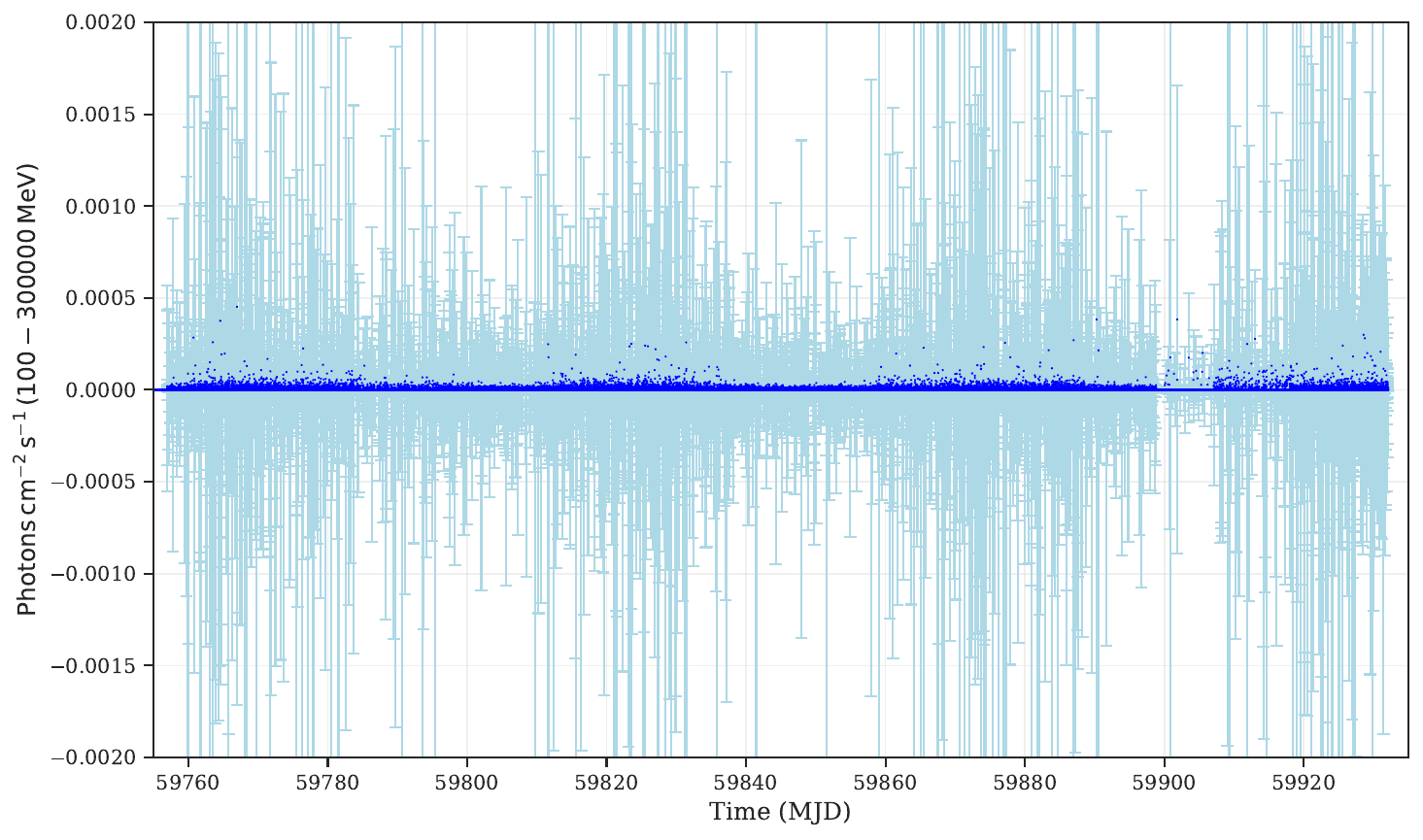}
\includegraphics[width=0.48\textwidth]{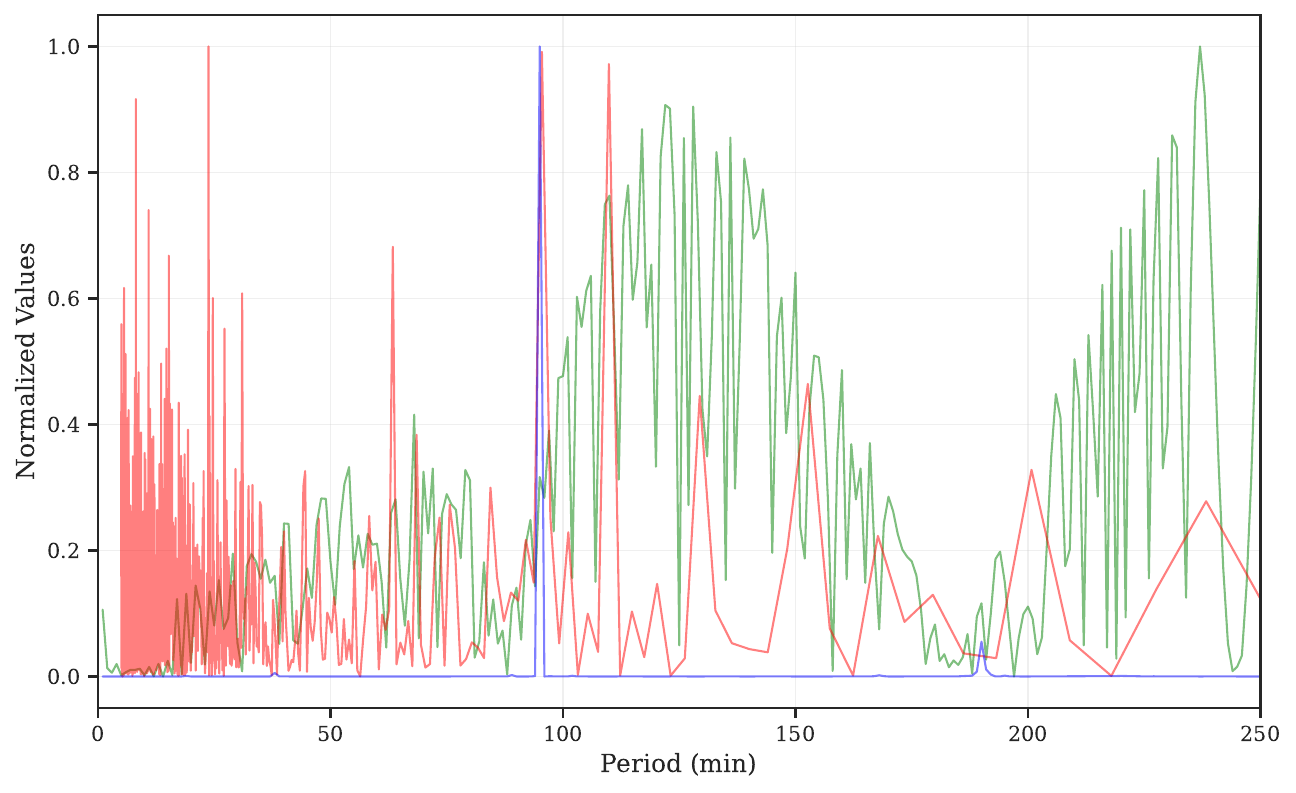}
\hfill
  \includegraphics[width=0.48\textwidth]{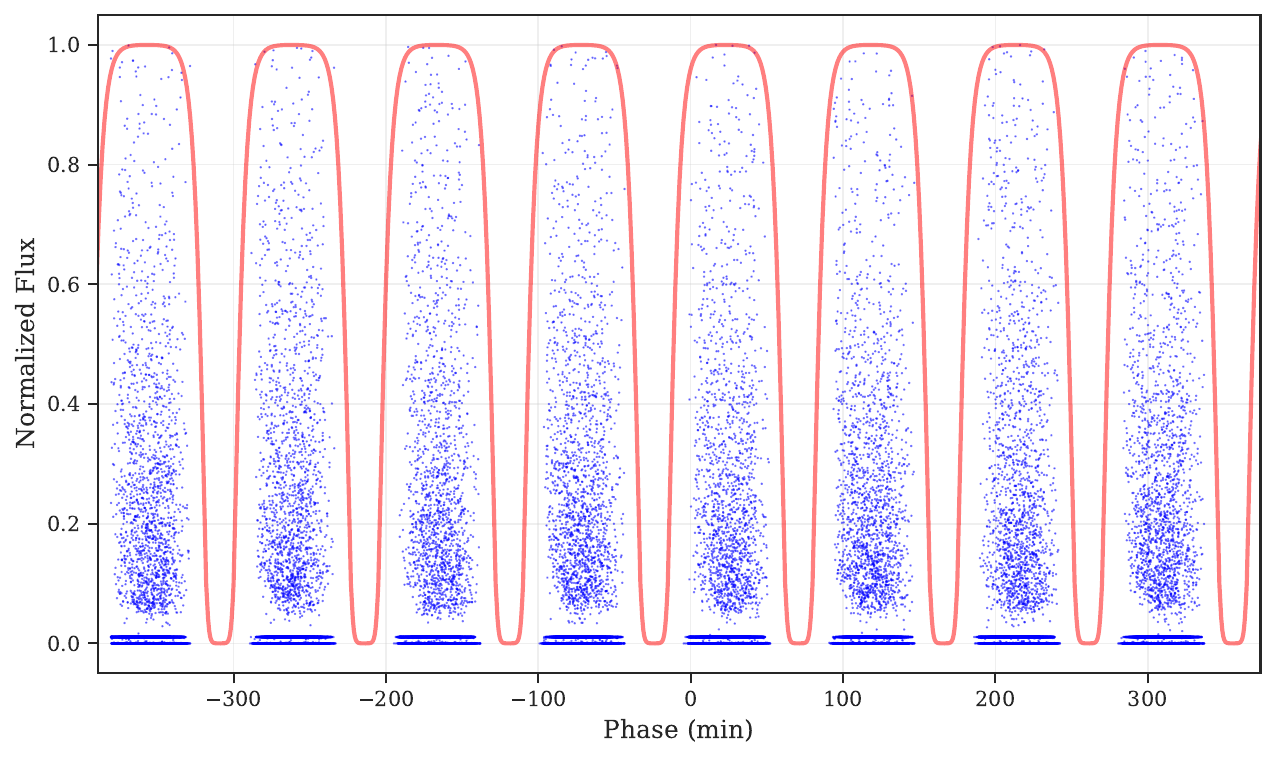}
 \caption{
The top panel shows the \(\gamma\)-rays light curve of Sgr~A* over 180
days from Dataset I (2008-08-04 to 2019-09-28) at a 3\(\sigma\) confidence
level (see Table \ref{table:table_1} for details). The bottom left
panel highlights the most significant peaks, with normalised values,
at 95 minutes using RobPer (blue line) and 95.47 minutes using the
Lomb-Scargle periodogram (red line). The Window Function (green curve)
value at 95 minutes is 0.316, confirming that the detected peak in the
periodograms is not significantly affected by sampling artifacts. The
bottom right panel displays the phase-folding of the light curve over
eight periods, using the parameters obtained from data fitting (see
Table~\ref{tab:table_2}) without error bars.}
\label{fig:figure_1}
 
\end{figure*} 
%%================================%%
%%====== E N D  F I G U R E ======%%
%%================================%%

Sgr~A* exhibits remarkable variability in emission across multiple
wavelengths. Initially detected in radio by \citet{Balick1974ApJ},
with subsequent observations extended to the infrared and X-rays 
bands~\citep{Boyce2019ApJ, Fazio2018}.

The detection of \(\gamma\)-rays from Sgr~A*~\citep{Cafardo2021ApJ}
highlights high-energy processes such as particle acceleration and
interactions, shedding light on extreme phenomena near the supermassive
black hole.

This letter presents our analysis of \(\gamma\)-rays variability using
three datasets. Dataset I comprises a continuous six-month time series
from 2022, while Datasets II and III include 519 and 347 light curves,
respectively, spanning 2008–2024 (see Table \ref{table:table_1}).  

%%===============================%%
%%==== B E G I N    T A B L E ===%%
%%===============================%%
\begin{table*}%[h!]
\centering
\begin{tabular}{cccccccc}
\hline
Dataset & Dates & Duration & Total & Sample & \(\alpha\) & \(\delta\) & \(r\) \\
        &       & (y, m, d) & (d)   & (d)    & (deg)      & (deg)      & (deg) \\
\hline
I   & 2022-06-22 -- 2022-12-19 & \(0, 5, 26\) & 180 & 180 & 266.41682600 & -29.00779700 & 15 \\
II   & 2008-08-04 -- 2019-09-28 & \(11, 1, 24\) & 4080 & 519 & 266.41681662 & -29.00782497 & 1 \\
III & 2019-09-28 -- 2024-10-28 & \(5, 1, 0\) & 1856 & 317 & 266.41682600 & -29.00779700 & 1 \\
\hline
\end{tabular}
\caption{The table shows the datasets, the time intervals, the
right ascension (\(\alpha\)), the declination (\(\delta\)), and the
radius (\(r\)) in Sgr~A*.}
\label{table:table_1}
\end{table*}
%%===============================%%
%%==== E N D        T A B L E ===%%
%%===============================%%

\section{Methods}
\label{methods}

\noindent
Using publicly available \(\gamma\)-rays
data from the Fermi satellite and the Fermitools
package\footnote{\url{https://fermi.gsfc.nasa.gov/ssc/data/analysis/software/}},
the light curves were constructed following the methods outlined by
\citet{gustavo, nacho}. The time was recorded in Modified Julian Days
(MJD), and the flux was measured in photons \(\mathrm{cm}^{-2}\)
\(\mathrm{s}^{-1}\)(100 -- 300000 MeV). A 3\(\sigma\) threshold was
applied to reduce noise and enhance the precision of metrics such as
means, variances, and the identification of periodicity patterns.

\subsection{RobPer Periodogram}

\noindent
The R-package RobPer facilitates the examination of periodograms
across extended time scales spanning several years. Specifically
designed for the analysis of astrophysical time series, as discussed 
by~\citet{anita02}, RobPer stands out due to its versatility in accommodating
various regression models, offering flexibility in data analysis. For
identifying critical values, the package employs an outlier detection
approach to validate periods. This method acknowledges the limitations
of traditional assumptions, aiming to identify periodicities in a more
robust and reliable manner. While sensitive to significant fluctuations,
RobPer provides a resilient framework for handling outliers effectively.

\subsection{Window Function}

\noindent
To improve the reliability of period detection in periodogram analysis
and reduce the likelihood of false positives, we employed a windowing
program as described by~\citet{dawson10}. This method effectively
differentiates between aliases and false positives by evaluating
whether the predicted aliases match the observed data in terms of
amplitude, phase, and pattern. By assessing the consistency of these
characteristics across all aliases, we can confidently determine the
presence of the true period. When all aliases align coherently, it
provides strong evidence that the genuine period has been 
identified~\citep{magallanes-mendoza-mrk501}.

\subsection{Lomb-Scargle periodogram and Bootstrap method}

\noindent
This algorithm, proposed by~\citet{Lomb1976Ap} and~\citet{Scargle1982},
is the most widely used method for detecting periodicities in unevenly
sampled light curves, integrating Fourier methods, phase-folding, least
squares, and Bayesian approaches~\citep{VanderPlas2018, baluev15}. A
bootstrap resampling procedure~\citep{efron1979bootstrap} was applied to
generate resampled subsets from the original data, enabling the
recalculation of periodograms and identification of maximum periods.
From these, a 95\% confidence interval was derived, capturing the data's
inherent variability and providing a reliable measure of the dominant
period's precision.

\subsection{\textit{K}-Means Clustering}

\noindent
Using the unsupervised machine learning algorithm
\textit{K}-Means~\citep{lloyd1982least, macqueen1967some,
forgy1965cluster}, clusters of significant periodicity values were
identified. The periodicities around 95~min were grouped (cluster) based
on the intrinsic characteristics of the data~\citep{scikit-learn}. This
approach enabled the establishment of a 95\% confidence interval for the
mean, allowing for an evaluation of the variability and reliability of
the average values within each cluster.

\subsection{Monte Carlo Markov Chains}

\noindent
The Monte Carlo Markov Chains (MCMC) method~\citep{brooks2011handbook}
was used to assess the significance of the detected periodic signals \(
p \)~\citep{sharma2017markov}. During each iteration of the simulation,
the flux data were randomly permuted while keeping the associated time
for each measurement constant. This approach generated a distribution of
maximum powers under the null hypothesis, which assumes no periodicity
in the data.

\subsection{Colour Analysis Noise}

\noindent
The Fourier transform~\citep{Aschwanden2011} was used to identify
periodic pulses in the light curves, even with excessive noise (see,
e.g., \citet{Press1978}). The power spectral density (PSD) reveals
periodic fluctuations with peaks at specific frequencies \(\nu\). The PSD
follows a power law, \( P(\nu) = \nu^{-\alpha} \), where  \( \alpha \)
is a fixed exponent. The spectral noise is characterised by its colour,
with white noise indicating no temporal correlation, Brownian noise
signifying significant temporal correlation~\citep{Schroeder1991},
and pink noise representing a phase change from random to predictive
behaviour~\citep{May1976, Aschwanden2010}. The power spectrum density
(PSD) exponent \(\alpha\) defines noise colours as follows: white (\(0.0
\lesssim \alpha \lesssim 0.5\)), pink (\(0.5 \lesssim \alpha \lesssim
1.5\)), and Brownian (\(1.5 \lesssim \alpha \lesssim 2.5\)).

\subsection{Data fitting with Jacobi elliptical function}

\noindent
The datasets were fitted with the Jacobi elliptical function using the
likelihood method. A parameter space search was conducted, exploring the
amplitude (A), Jacobi modulus (m), enhancement period (E), and silence
period (S). One million simulations were performed across all the light
curves to identify the optimal parameters that maximise the coherence
between the simulations and the observations. The parameters were
adjusted to the data using the Jacobi elliptical function fitting technique
described by \citet{magallanes-mendoza-mrk501}.

\subsection{Phase folding}
\label{sec:phasefold}

\noindent
Once the period was determined, the data were plotted without error bars
for greater clarity, as a function of phase rather than time. This
allows multiple cycles of the phenomenon to be aligned in a single 
graph~\citep{Schwarzenberg1997ApJ, Linnell1985AJ}. The success of phase-folding largely depends on the accuracy of the period determination. For
this reason, prior analysis of the period using statistical and
computational methods must be considered to ensure effective 
phase-folding~\citep{Stella1998ApJ}.

\section{Results}
\label{results}

\noindent
Using Dataset I, we constructed the light curve (Figure
\ref{fig:figure_1} and Table \ref{table:table_1}) and identified
significant periodicities using the RobPer periodogram. The maximum
peaks for RobPer and Lomb-Scargle were at 95 and 95.47 minutes
respectively, with a WF value of 0.031, confirming the peak was not
influenced by sampling artifacts. Parameters were fitted using the Jacobi
elliptical function. Results with the confidence interval are shown in
Table~\ref{tab:table_2}, and phase-folding was performed on
the light curve (Figure~\ref{fig:figure_1} bottom right).
 
\begin{figure*}%[h!]
\begin{center}
\includegraphics[width=0.48\textwidth]{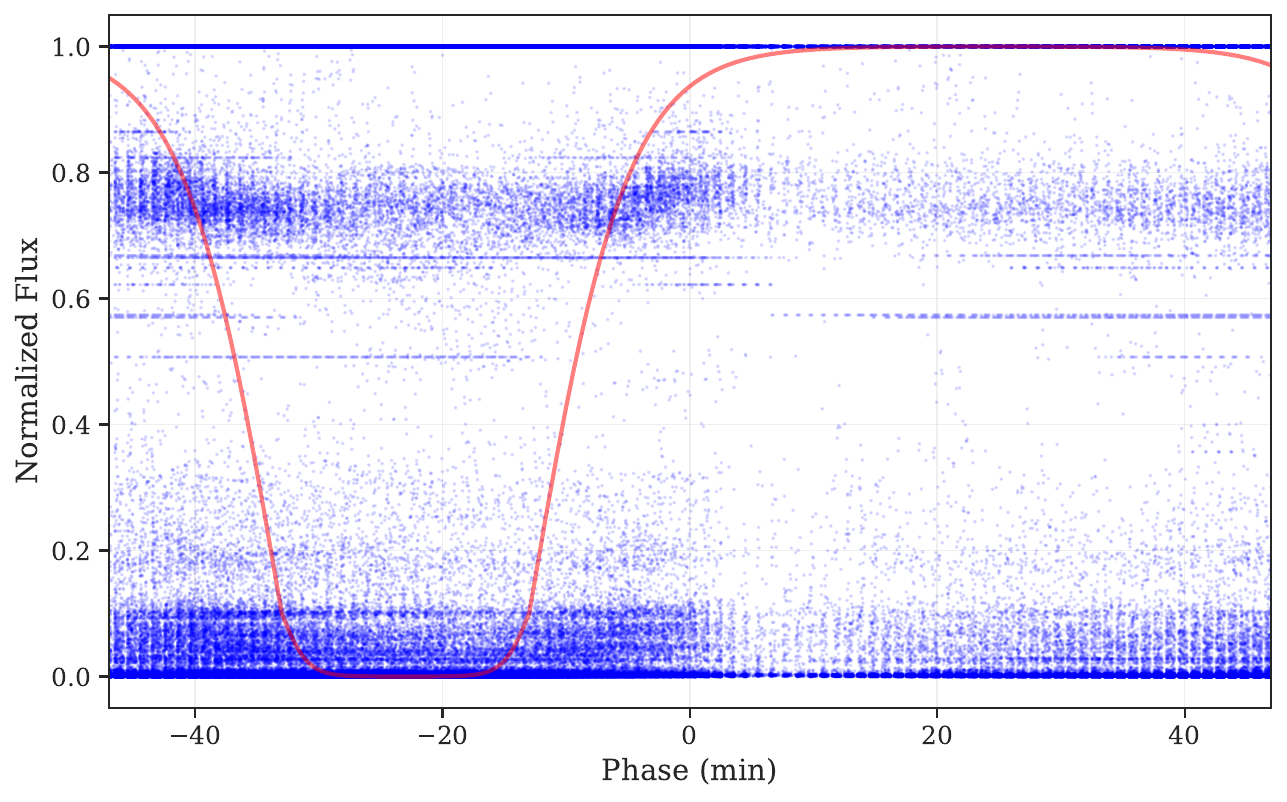}
\includegraphics[width=0.48\textwidth]{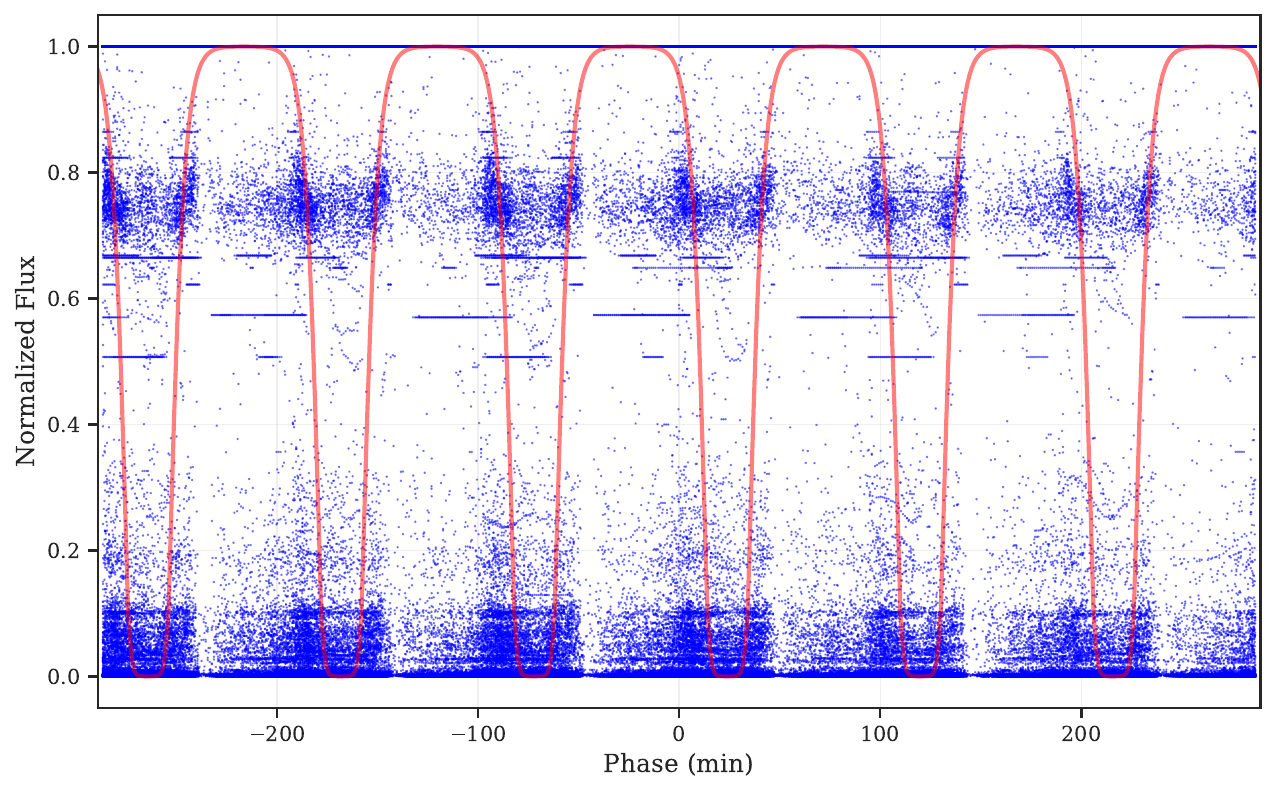}
\hfill
\includegraphics[width=1\textwidth]{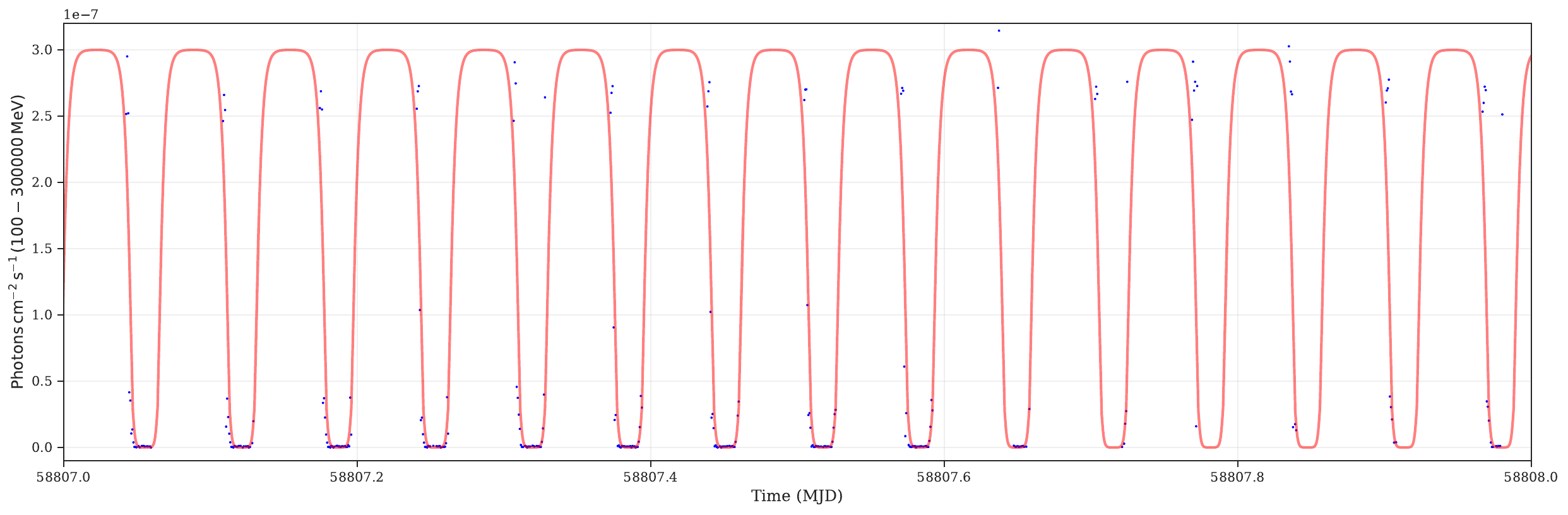}
\caption{The two top figures display the phase-folding of 519 normalised light
curves of Sgr~A* from a sample of Dataset II, fitted using the Jacobi
elliptical function with the likelihood method (see Table \ref{tab:table_2}). The left panel shows the
phase-folding over one period, while the next panel shows it over six
periods. The bottom figure shows the light
curve of 2022-01-17 00:00:04.000 UTC. The plots reveal a periodicity of
\(\sim\)~78 minutes, with a silent time of \(\sim\)~18 minutes, resulting in a total quasi-periodic oscillation (QPO) of around 96 minutes.}
\label{fig:figure_2}

\end{center} 
\end{figure*}

\begin{figure*}%[h!]
    \centering
    \includegraphics[width=0.48\textwidth]{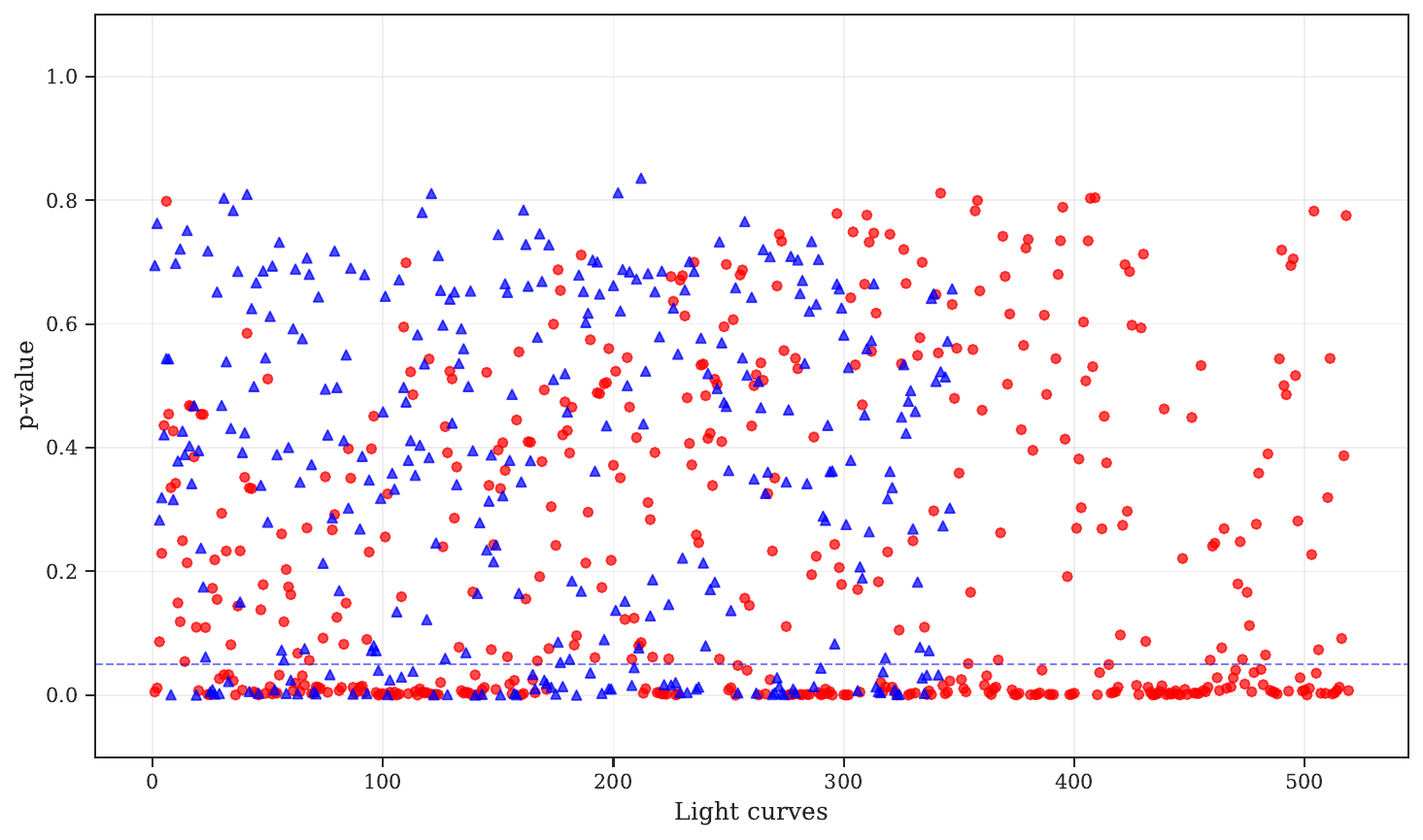}
    \includegraphics[width=0.48\textwidth]{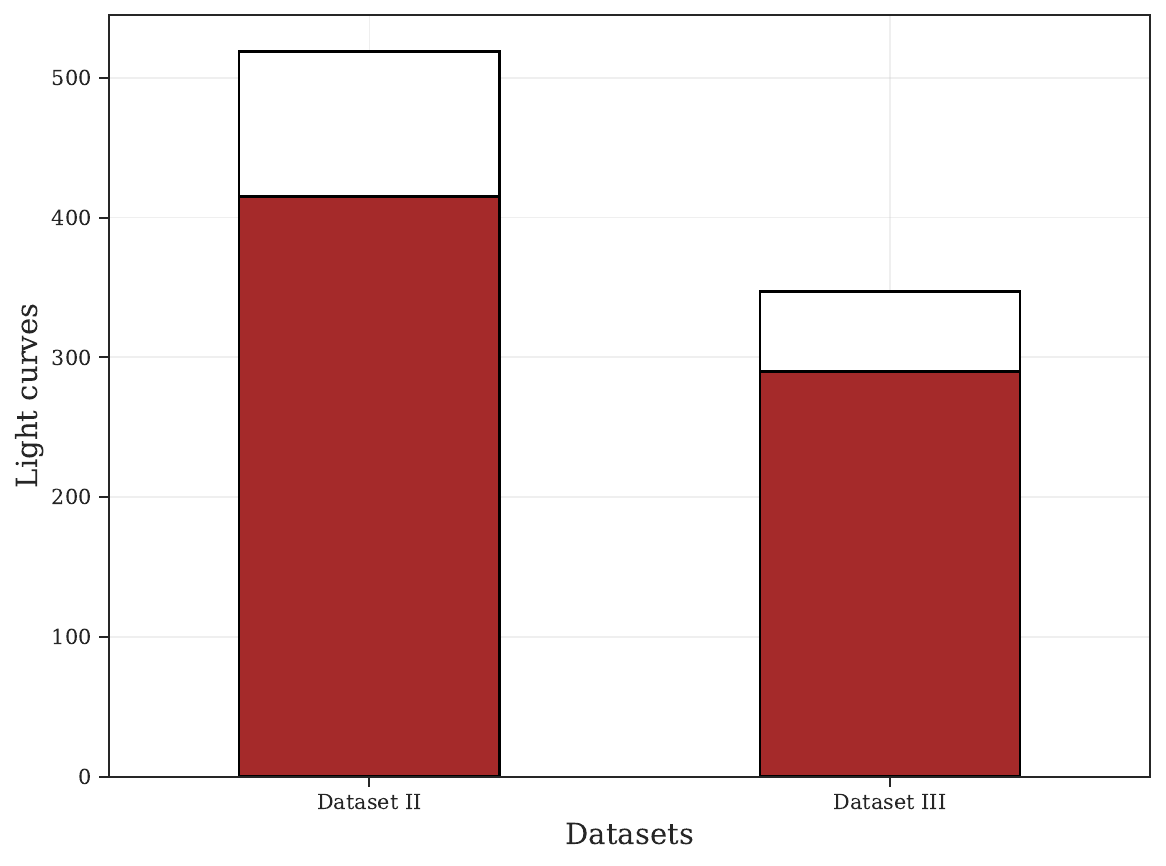}
    \caption{The left panel shows the significance levels from a Monte
Carlo simulation, with the dashed line indicating \(p = 0.05\). In
Dataset II (red circles), 61.27\% of the series exhibit significant
periodicity (\(p \leq 0.05\)) near 96 minutes, suggesting a genuine
periodicity. Similarly, in Dataset III (blue triangles), 79.25\% of the
analysed series also demonstrate significant periodicity. The right
panel focuses on noise colour distribution, revealing that 79.96\% of
Dataset II consists of Brownian noise (brown colour bar), and 83.57\% in Dataset III.}

    \label{fig:figure_3}

\end{figure*}

In Dataset II (Table~\ref{table:table_1}), we analysed 519 light curves
(see \url{https://www.guijongustavo.org/sgra/light_curves.html} for more
detail) using clustering (\(K\)-Means) and the RobPer
algorithm, identifying a period of 95.47 with I.C. (95.10–95.85).
Similarly, clustering with Lomb-Scargle revealed a period of
95.42 with I.C. (95.34–95.50). MCMC simulations confirmed
significant periodicity (\(p \leq 0.05\)) in 318 light curves
(61.27\%), while 79.96\% showed non-white noise (Brownian), strengthening
reliability (Figure \ref{fig:figure_3}). Parameter
space exploration via Jacobi elliptical function fitting is
detailed in Table~\ref{tab:table_2} and Figure
\ref{fig:figure_2}.

%%===============================%%
%%==== B E G I N  F I G U R E ===%%
%%===============================%%
%%===============================%%
%%====== E N D F I G U R E ======%%
%%===============================%%

%%===============================%%
%%==== B E G I N    T A B L E ===%%
%%===============================%%
\begin{table*}[!htbp]
    \centering
    \resizebox{\textwidth}{!}{
        \begin{tabular}{ccccccc}
            \hline
            \textbf{Parameters} & \multicolumn{2}{c}{\textbf{Dataset I}} & \multicolumn{2}{c}{\textbf{Dataset II}} & \multicolumn{2}{c}{\textbf{Dataset III}} \\
            \hline
            & \textbf{Value} & \textbf{C. I.} & \textbf{Value} & \textbf{C. I.} & \textbf{Value} & \textbf{C. I.} \\
            \hline
            A & 25.443166 & (20.300561, 27.790759) & 28.708196 & (22.648931, 34.354090) & 30.648994 & (24.278314, 36.157631) \\
            m & 0.990170  & (0.990091, 0.990711)  & 0.992737  & (0.992112, 0.993788)  & 0.995331  & (0.992206, 0.997832) \\
            E & 79.076802 & (78.966522, 79.277777) & 78.009511 & (75.697586, 79.569630) & 78.628943 & (76.851035, 79.534460) \\
            S & 18.129304 & (18.015291, 18.300728) & 18.684087 & (18.076002, 19.577166) & 18.274335 & (18.079911, 18.730829) \\
            \hline
        \end{tabular}
    }
    \caption{Parameters obtained with the likelihood method using Jacobi
elliptical functions for datasets I, II, and III, with 95\% Confidence Intervals.}
    \label{tab:table_2}

\end{table*}
%%===============================%%
%%==== E N D        T A B L E ===%%
%%===============================%%

For Dataset III (Table \ref{table:table_1}), we analysed 347
light curves, detecting periodicities
with RobPer at 94.98 (I.C. 94.89–95.07) and Lomb-Scargle at
95.28 (I.C. 95.20–95.37) using \(K\)-Means clustering. MCMC
simulations identified significant periodicity (\(p \leq 0.05\))
in 275 light curves (79.25\%), with a 96-minute period, confirming
its presence (Figure \ref{fig:figure_3}). Moreover,
83.57\% of light curves exhibited non-white noise (Brownian), reinforcing
validity. Parameter exploration through Jacobi elliptical function
fitting is detailed in Table~\ref{tab:table_2} and
Figure~\ref{fig:figure_2}. These results from the noise colour analysis
and the MCMC also correspond to Dataset I.

%%===============================%%
%%==== B E G I N  F I G U R E ===%%
%%===============================%%
%%=================================%%
%%====== E N D   F I G U R E ======%%
%%=================================%%

\section{Discussion}

\label{discussion}
For Dataset I (with a 15-degree radius), RobPer, Lomb-Scargle, and Window
Function algorithms identified a \(\sim\)96-minute period. Data fitting
with the Elliptical Jacobi Function and phase folding further validated
this period. For the larger Datasets II and III, we used sampling,
bootstrap, \(K\)-Means clustering, and noise colour analysis within
a 1-degree radius, consistently confirming this periodicity with robust
methodologies.

Data fitting across the three datasets was performed
using the Jacobi elliptical functions method described by
\citet{magallanes-mendoza-mrk501}.
This approach calculates
oscillation coordinates via the elliptical function \(\mathrm{sn}\),
dependent on a modulus \(m\) controlling curvature [see e.g.,
\citep{narayan1996lectures}], the amplitude (A), the enhancing (E)
and silence (S) periods.

  This time-dependent oscillation is modulated by elliptical functions and
periods E and S. Finally, likelihood values were calculated to evaluate
how well the parameters explained the data, with 1,000,000 iterations
performed across all parameter combinations to assess their performance.

\section{Conclusions}
\label{conclusions}

\noindent
Using statistical and computational methods, the 96-minute
quasi-periodicity was confirmed. This periodicity was found performing
a weighted estimation, assigning relative weights to values based
on their Confidence Intervals. This yielded an enhancement of 78.57
(CI: 77.17–79.46), a silence time of 18.36 (CI: 18.05–18.86), an
amplitude of 28.26 (CI: 22.40–32.76), and a Jacobi module of 0.992 (CI:
0.991–0.994). With this we confirm the \(\gamma-rays\) QPO intraday
of 96 min in Sgr~A*.

The origin of this periodicity remains unclear and is beyond
the scope of this letter. The periodicity \(\sim\)78 min is closely related
to the one detected in the radio waveband by~\citet{Wielgus2022} and 
approximates half of the one found in X-rays by~\citet{Leibowitz2018MNRAS}. All these periodicities likely stem
from the same mechanism described by~\citet{Wielgus2022}: a magnetised
matter blob orbiting the supermassive black hole in Sgr~A*. However,
high-energy processes such as particle acceleration and interactions
like \(\gamma\)~-rays shed light on the extreme phenomena near the
supermassive black hole.

\section*{Acknowledgements}
\label{acknowledgements}
% We thank an anonymous referee for the very useful comments 
% made to the manuscript.

This work was supported by DGAPA-UNAM (IN110522 and IN118325). 
GMG and SM acknowledge economic support from Secihti
(378460, 26344). 

We thank the Fermi Gamma-Ray Space Telescope collaboration for providing
the public data used in this work.

%%%%%%%%%%%%%%%%%%%%%%%%%%%%%%%%%%%%%

\section*{Data Availability}
The data underlying this article will be shared on 
request to the authors.

%%%%REFERENCES:
\bibliographystyle{mnras}
\bibliography{magallanes-mendoza}

%%%%%%%%%%%%%%%%%%%%%%%%%%%%%%%%%%%%%%%%%%%%%%%%%%

% Don't change these lines
\bsp	% typesetting comment
\label{lastpage}
\end{document}